\documentstyle[12pt]{article}
 
\begin{document}

\hspace*{1ex}

\mbox{\hspace*{58ex}} OCU-PHYS-167 

\mbox{\hspace*{57ex}} October 1997 

\vspace*{10mm}

\begin{center}
{\Large {\bf 
Perturbative QCD at finite temperature and density\footnote{Invited 
talk presented at International School on the Physics of Quark Gluon 
Plasma, June 3-6, Hiroshima} }} 
\end{center}

\hspace*{3ex}

\hspace*{3ex}

\hspace*{3ex}

\begin{center} 
{\large {\sc A. Ni\'{e}gawa}\footnote{ 
E-mail: niegawa@sci.osaka-cu.ac.jp}

{\normalsize\em Department of Physics, Osaka City University } \\ 
{\normalsize\em Sumiyoshi-ku, Osaka 558, Japan} } \\
\end{center} 

\hspace*{2ex}

\hspace*{2ex}

\hspace*{2ex}

\hspace*{2ex}
\begin{center} 
{\large {\bf Abstract}} \\ 
\end{center} 
\begin{quotation}
This is a comprehensive review on the perturbative hot QCD 
including the recent developments. The main body of the review is 
concentrated upon dealing with physical quantities like reaction 
rates. 

Contents: 
\S1. Introduction, 
\S2. Perturbative thermal field theory: Feynman rules, 
\S3. Reaction-rate formula, 
\S4. Hard-thermal-loop resummation scheme in hot QCD, 
\S5. Effective action, 
\S6. Hard modes with $|P^2| \leq O (g^2 T^2)$, 
\S7. Application to the computation of physical quantities, 
\S8. Beyond the hard-thermal-loop resummation scheme, 
\S9. Conclusions. 
\end{quotation}
\newpage
\setcounter{equation}{0}
\setcounter{section}{0}
\def\theequation{\mbox{\arabic{section}.\arabic{equation}}}
\section{Introduction}
Among various predictions of QCD, a distinguished one is the 
existence of the QCD phase transition \cite{sama} at the temperature 
$T_c \simeq 150 MeV$. The phase at the high-temperature side, 
$T > T_c$, is called the quark-gluon plasma (QGP) phase. According 
to the standard big bang scenario, the QGP has lived in the early 
days of the Universe. Efforts of reviving the QGP in the present day 
are making and it is expected to be realized soon. 

Theoretical approaches to the QGP physics fall into four main 
categories; lattice simulation, perturbative approach, 
effective-theory approach and phenomenological approach. These 
approaches are complementary to each other and each of them has its 
own advantage. 

The purpose of this paper is to give a comprehensive review of the 
perturbative approach. \cite{sama,kapu,ume,das,chou,land,henn} 
Thanks to the asymptotic freedom of QCD the coupling constant $g$ 
decreases with temperature $T$ and/or density and then at high 
temperature/density the perturbative approach comes to be an 
powerful device. 

Thermodynamic properties of a QGP system in thermal and chemical 
equilibrium is characterized by 
\begin{eqnarray*} 
Z   & = & 
        \mbox{Tr} \, e^{- \beta (H - \mu Q)} 
        \;\;\;\;\; (\beta = 1 / T) \, , \\ 
        \ln Z & = & P V / T \, , 
\end{eqnarray*} 
where $Z$, $H$, $P$ and $V$ are, in respective order, the 
grand-partition function, QCD Hamiltonian, pressure and volume of 
the system, and $\mu$ is the chemical potential being conjugate to 
the quark number $Q$. 
[For a concise review of the properties of a QGP, I refer to 
Ref. 8).] The thermal average of a quantity $\Omega$ is defined by 
\[ 
\langle \Omega \rangle \equiv \mbox{Tr} \, \Omega \, 
e^{- \beta (H - \mu Q)} / Z \, . 
\] 
Rates of reactions taking place in a QGP are computed through 
thermal Green functions, which are defined as the thermal average of 
the relevant products of field operators. 
\setcounter{equation}{0}
\setcounter{section}{1}
\section{Perturbative thermal field theory: 
\cite{sama,kapu,ume,das,chou,land} Feynman rules} 
\def\theequation{\mbox{\arabic{section}.\arabic{equation}}}
Traditional approach starts with taking in-fields in vacuum ($T = 
0$) theory as a basis of a Fock space. 
\subsection{Imaginary-time or Matsubara formalism} 
This formalism is convenient for calculating no-leg amplitudes 
(free energy, grand-partition function etc.) and two-point 
functions. 

$\bullet$ Propagator: 
\[ 
\frac{{\cal N} (P)}{p_0^2 + {\bf p}^2 + m^2} \, . 
\] 
Here $P = (p_0, {\bf p})$ with $p_0 = 2 \pi n T \, [\pi (2 n + 1) T 
- i \mu]$ for gluon or FP ghost [quark] ($n = \cdot \cdot \cdot, -2, 
-1, 0, 1, 2, \cdot \cdot \cdot)$. The form of the \lq\lq numerator 
factor'' ${\cal N} (P)$ is the same as in Euclidean vacuum theory. 

$\bullet$ Vertex: To an $N$-particle vertex, is assigned 
\[ 
{\cal V}_{...} \, (2 \pi)^3 \, \frac{1}{T} \, \delta_{n, 0} \, 
\delta ({\bf p}) \, , 
\] 
where $n = \sum_{i = 1}^N n_i$, ${\bf p} = \sum_{i = 1}^N {\bf p}_i$ 
and ${\cal V}_{...}$ stands for the factor that can be read off from 
the interaction Lagrangian. 

$\bullet$ Internal momentum: 
\[ 
T \sum_{n} \int \frac{d^{\, 3} p}{(2 \pi)^3} \, . 
\] 
\subsection{Real-time formalism} 
This formalism allows us to directly compute $N$-point functions. 
The theory is formulated by introducing a contour $C = C_1 + C_2 + 
C_3 + C_4$ in a complex time plane: $t_i \to t_f$ ($C_1$), $t_f 
\to t_f - i \sigma$ ($C_3$), $t_f - i \sigma \to t_i - i \sigma$ 
($C_2$) and $t_i - i \sigma \to t_i - i \beta$ ($C_4$), where 
$0 \leq \sigma \leq \beta$. Then, the limit $t_{i / f} \to \mp 
\infty$ is taken. As far as thermal amplitudes are concerned, the 
contributions from the contour segments $C_3$ and $C_4$ may, {\em in 
a sense}, be ignored. \cite{nie,sama,land} Thus the formalism turns 
out to be a two-component theory; the field whose time argument is 
in the segment $C_1$ ($C_2$), $\phi_1$ ($\phi_2$), is called the 
type-1 (type-2) or physical (thermal-ghost) field. Then, in this 
formalism, propagators, vertices and self-energy parts enjoy 
$2 \times 2$ matrix structure. Theories with different $\sigma$'s 
constitute an equivalent class of theories. Physical-field 
amplitudes are independent of $\sigma$. 

$\bullet$ Propagator: \cite{land} 
\begin{equation} 
i \, {\cal N} (P) \, \hat{\Delta} (P) \, . 
\label{rea} 
\end{equation} 
Here $P = (p_0, {\bf p})$ with $p_0$ real and [upper (lower) suffix 
refers to gluon and FP ghost (quark)] 
\begin{eqnarray} 
& & \hat{\Delta} (P) = 
      \hat{M}_\pm (P) \, \hat{\Delta}_F (P) \, 
      \hat{M}_\pm (P) \, , 
\label{real-pro1} \\ 
& & \hat{M}_\pm (P) = 
        \left( 
           \begin{array}{cc} 
             \sqrt{1 \pm n_\pm (p_0)} & e^{\sigma p_0} 
                 \frac{\theta (- p_0) \pm n_\pm 
                 (p_0)}{\sqrt{1 \pm n_\pm (p_0)}} \\ 
             e^{- \sigma p_0} \frac{\theta (p_0) \pm n_\pm 
             (p_0)}{\sqrt{1 \pm n_\pm (p_0)}} & \sqrt{1 \pm 
             n_\pm (p_0)} 
           \end{array} 
        \right) \, , 
\label{real-pro2} \\ 
& & \hat{\Delta}_F (P) = 
      \mbox{diag} 
         \left[ 
            \Delta_F (P) , \, - \Delta^*_F (P) 
         \right] 
            \;\;\;\;\;\;\;\;\; 
          \left( 
              \Delta_F (P) = 1 / (P^2 - m^2 + i 0^+) 
          \right) \, , 
\label{aaa} \\ 
& & \mbox{\hspace*{3ex}} n_+ (p_0) = 
        \frac{1}{e^{\beta |p_0|} - 1} \, , \;\;\;\;\;\;\; 
    n_- (p_0) = 
        \frac{1}{e^{\beta (|p_0| - \epsilon (p_0) \mu)} + 1} \, . 
\label{n} 
\end{eqnarray} 
$i \Delta_{1 1}$, $i \Delta_{1 2}$, $i \Delta_{2 1}$ and $i 
\Delta_{2 2}$ are Fourier transforms with respect to $(\mbox{Re} \, 
(x_0 - y_0), {\bf x} - {\bf y})$ of, in respective order, $\langle 
T \phi (x) \overline{\phi} (y)\rangle$, $\tau \, \langle 
\overline{\phi} (y) \phi (x) \rangle$, $\langle \phi (x) 
\overline{\phi} (y)\rangle$ and $\langle \overline{T} \phi (x) 
\overline{\phi} (y)\rangle$, where $\overline{\phi}$ is the adjoint 
of $\phi$, $\tau = + (-)$ for gluon and FP-ghost (quark) and 
$\overline{T}$ is the anti-time-ordering symbol. 

$\bullet$ Vertex 
\[ 
i \, {\cal V}_{...} (2 \pi)^4 \, \delta^{\, 4} (P) 
    \left( 
       \begin{array}{cc} 
          1 & 0 \\ 
          0 & - 1 
       \end{array} 
    \right) \, , 
\] 
where $P = \sum_{i =1}^N P_i$. Note that the vertex matrix is 
diagonal. The $(1, 1)$ component is called the type-1 vertex, which 
is same as the vacuum-theory counterpart, and the $(2, 2)$ component 
is called the type-2 vertex. 

$\bullet$ Internal momentum: 
\[ 
\int \frac{d^{\, 4} P}{(2 \pi)^4} \, . 
\] 

\hspace*{1ex} 

\begin{center} 
Notes: 
\end{center} 

1) Due to thermal radiative corrections, full gluon- and 
quark-propagators split into several pieces. However their 
structure remains \cite{land} essentially to be the same as the 
corresponding bare propagator, Eqs. (\ref{real-pro1})~-~(\ref{n}) 
(cf. below). 
Incidentally, (each piece of) the self-energy part takes the form 
\begin{equation} 
\hat{\Sigma} (P) = 
   \hat{M}_\pm^{- 1} (P) 
      \left( 
         \begin{array}{cc} 
            \Sigma_F (P) & 0 \\ 
            0 & - \Sigma^*_F (P) 
         \end{array} 
      \right) 
   \hat{M}_\pm^{- 1} (P) \, . 
\label{self} 
\end{equation} 

2) Thermo field dynamics \cite{ume,das,land} (A two-component theory 
formulated with canonical quantization): For each field $\phi (x)$ 
in the original Hamiltonian, introduce its \lq\lq copy'' 
$\tilde{\phi} (x)$. $\tilde{\phi}$ is essentially the field obtained 
from $\phi$ through time-reversal operation. [$\phi$ 
($\tilde{\phi}$) corresponds to $\phi_1$ ($\phi_2$) in the time-path 
ordered formalism outlined above. Note that $\phi_2$ is in the 
contour segment $C_2$, in which Re $t$ flies backward, $+ \infty \to 
- \infty$.] Then, a quasiparticle field $\varphi_i$ $(i = 1, 2)$ is 
introduced, in momentum space, through 
\begin{eqnarray} 
& \hat{\varphi} (P) \equiv 
    \hat{M}_\tau^{- 1} (P) \hat{\phi} (P) 
         \;\;\;\;\;\;\;\;\;\;\;\;\;\; & \hat{\phi} (P) = 
           \left( 
              \begin{array}{c} 
                  \phi (P) \\ 
                  \tilde{\phi} (P) 
              \end{array} 
           \right) \, , \nonumber \\ 
& \hat{\overline{\varphi}} (P)  = 
    \hat{\overline{\phi}} (P) \, \hat{M}_\tau^{- 1} (P) \, , 
    \;\;\;\;\;\;\;\;\;\;\; 
    & \hat{\overline{\phi}} (P) = 
        (\overline{\phi} (P), \overline{\tilde{\phi}} (P)) \, , 
\label{TFD} 
\end{eqnarray} 
where $\tau = + [-]$ for gluon and FP ghost [quark]. Note that, in 
general, $\overline{\varphi}_i$ is not the adjoint of $\varphi_i$. 
The so-called thermal vacuum is introduced by $\varphi_i^{(\xi_i)} 
(x) | 0 \rangle = \langle 0 | \overline{\varphi}_i^{(- \xi_i)} (x) 
= 0$ $(\xi_1 = +, \, \xi_2 = -)$, where \lq\lq $(+) / (-)$'' 
indicates the positive/negative frequency part. With this 
preliminaries, we see that $\langle 0 | \, T \hat{\phi} \, 
\hat{\overline{\phi}} \, | 0 \rangle = i {\cal N} \hat{\Delta}$, Eq. 
(\ref{rea}). Thus, as far as the perturbation scheme is concerned, 
both thermo field dynamics and the time-path formalism summarized 
above are equivalent. 
\setcounter{equation}{0}
\setcounter{section}{2}
\section{Reaction-rate formula \cite{nie1,nie11}} 
\def\theequation{\mbox{\arabic{section}.\arabic{equation}}}
To avoid inessential complications, I take a heat bath composed of 
massless scalar fields. Reactions taking place in the heat bath are 
of the following generic type; 
\begin{equation} 
\Phi (P_1) + \cdot \cdot \cdot + \Phi (P_n) + \mbox{heat bath} 
    \to 
       \Phi (Q_1) + \cdot \cdot \cdot + \Phi (Q_m) + \mbox{anything} 
       \, . 
\label{reaction} 
\end{equation} 
Here $\Phi$ is a nonthermalized heavy scalar particle. 
[Generalization to other cases is straightforward.] 

The reaction rate $R$ reads 
\begin{eqnarray} 
& & \frac{1}{V} \left( \prod_{j = 1}^m 2 q_{j 0} \frac{d}{d 
   {\bf q}_j / (2 \pi)^3} \right) R \nonumber \\ 
& & \mbox{\hspace*{3ex}} = 
   \left( 
      \prod_{i = 1}^n \frac{1}{2 p_{i 0} V} 
   \right) 
   A 
   \left( 
      P_1^{(2)}, \cdot \cdot \cdot, P_n^{(2)}, Q_1^{(1)}, 
      \cdot \cdot \cdot, Q_m^{(1)}; P_1^{(1)}, \cdot \cdot \cdot, 
      P_n^{(1)}, Q_1^{(2)}, \cdot \cdot \cdot, Q_m^{(2)} 
   \right) 
     \, , \nonumber \\ 
\label{rate} 
\end{eqnarray} 
where $p_{i 0} = E_i = \sqrt{p_i^2 + m^2}$ etc. In Eq. (\ref{rate}), 
$A$ is an amplitude evaluated in the Keldish variant ($\sigma = 0$ 
in \S~2.2) of real-time formalism for the \lq\lq process,'' 
\begin{eqnarray*} 
& & \Phi_1 (P_1) + \cdot \cdot \cdot + \Phi_1 (P_n) + \Phi_2 (Q_1) + 
   \cdot \cdot \cdot + \Phi_2 (Q_m) \nonumber \\ 
& & \mbox{\hspace*{5ex}} \to \Phi_2 (P_1) + \cdot \cdot \cdot + 
   \Phi_2 (P_n) + \Phi_1 (Q_1) + \cdot \cdot \cdot + \Phi_1 (Q_m) 
   \, , 
\end{eqnarray*} 
where the suffix \lq $i$' ($i = 1, 2$) refers to type-$i$ field. 

\hspace*{1ex} 

\begin{center} 
Addenda: 
\end{center} 

1) 
$A$ in Eq. (\ref{rate}) is {\em not} an absolute square of some 
amplitude, in contrast to the case of vacuum theory. 

2) 
When $\Phi (P_i)$ [$\Phi (Q_j)$] in the reaction (\ref{reaction}) is 
a thermalized particle, the factor $n_B (E_i)$ [$1 + n_B (E_j)$] is 
to be multiplied to the right-hand side (rhs) of Eq. (\ref{rate}). 
Here $n_B$ is the Bose distribution function. 

3) 
The formula (\ref{rate}) is valid \cite{nie11} even for a finite 
cube system as far as the periodic boundary condition is employed 
for the single-particle wave function basis. 

4) 
In the limit $T \to 0$, the formula (\ref{rate}) reduces 
\cite{nie11} to the formula obtained through Cutkosky or cutting 
rules. In particular, for $n = 2$ and $m = 0$, the formula goes 
to the optical theorem and, for $n = 2$ and $m = 1$, the formula 
goes to the Mueller formula for the corresponding inclusive 
reaction. 

5) 
Applying the formula (\ref{rate}) to the reaction (\ref{reaction}), 
where $\Phi$'s are constituent particles of the heat bath (cf. the 
second item above), one can derive \cite{nie11} the detailed-balance 
formula. Namely, the rate (\ref{rate}) is equal to the rate for 
the inverse process to (\ref{reaction}). 

6) 
Cutting rules: The reaction-rate formula (\ref{rate}) is derived 
from the \lq\lq first-prin\-ci\-ple formula'' 
\begin{equation} 
R \propto \mbox{Tr} \, e^{- \beta (H - \mu Q) } S^* S / \mbox{Tr} \, 
   e^{- \beta (H - \mu Q)} \, , 
\label{start} 
\end{equation} 
where $S$ is the $S$-matrix element in {\em vacuum theory} for the 
process 
\[ 
\Phi (P_1) + \cdot \cdot \cdot + \Phi (P_n) + \{ \phi\mbox{'s} \} 
    \to 
        \Phi (Q_1) + \cdot \cdot \cdot + \Phi (Q_m) + 
        \{ \phi\mbox{'s} \} \, . 
\] 
Here $\phi$'s are constituent particles of the heat bath. 

In what follows, we depart from the scalar theory and keep in mind 
some general theory. The $(1, 1)$ component of the thermal 
propagator $\Delta_{1 1} (P)$ in $A$, Eq. (\ref{rate}), has the 
following three roots; (i) the $T = 0$ propagator connecting the 
vertices $v_1$ and $v_2$ in $S$, Eq. (\ref{start}), which carry the 
momentum $P$ from $v_1$ to $v_2$, (ii) a particle of momentum 
${\bf p}$ is absorbed from the heat bath into the vertex $v_2$ and a 
particle of the same momentum ${\bf p}$ is emitted from the vertex 
$v_1$ into the heat bath and (iii) an antiparticle of momentum 
$- {\bf p}$ is absorbed from the heat bath into the vertex $v_1$ and 
an antiparticle of the same momentum $- {\bf p}$ is emitted from the 
vertex $v_2$ into the heat bath. Both end-point vertices of 
$\Delta_{1 1} (P)$ are of type-1, which come from the vertices in 
$S$ in Eq. (\ref{start}). 

The roots of the $(2, 2)$ component of the thermal propagator 
$\Delta_{2 2} (P)$ in $A$ are obtained from above by $S \to S^*$. 
Both end-point vertices of $\Delta_{2 2} (P)$ are of type-2, which 
come from the vertices in $S^*$ in Eq. (\ref{start}). 

The $(2, 1)$ component of the thermal propagator $\Delta_{2 1} (P)$ 
in $A$ has the two roots; (i) a particle of momentum ${\bf p}$ is 
emitted into the heat bath from a vertex $v_1$ in $S$, Eq. 
(\ref{start}), and a particle of the same momentum ${\bf p}$ is 
absorbed from the heat bath into a vertex $v_2$ in $S^*$, Eq. 
(\ref{start}), and (ii) an antiparticle of momentum $- {\bf p}$ is 
absorbed from the heat bath into $v_1$ and an antiparticle of the 
same momentum $- {\bf p}$ is emitted from $v_2$ into the heat bath. 
The vertex $v_1$ ($v_2$) in $S$ ($S^*$) goes to the type-1 (type-2) 
end-point vertex of $\Delta_{2 1} (P)$. 

The roots of the $(1, 2)$ component of the thermal propagator 
$\Delta_{1 2} (P)$ are obtained from those for $\Delta_{2 1} (P)$ by 
$S \leftrightarrow S^*$. 

\hspace*{1ex} 

This inspection leads us to introduce the thermal cutting rules: 
\cite{nie1} Cut all the lines $\Delta_{1 2}$'s, $\Delta_{2 1}$'s, 
$\Delta_{1 1}^{(T)}$'s, and $\Delta_{2 2}^{(T)}$'s. [The superscript 
\lq\lq $(T)$'' refers to the $T$-de\-pend\-ent part.] 

\hspace*{1ex} 

Through an application of the above cutting rules to some 
reaction-rate formula, $A$ in Eq. (\ref{rate}) is divided into 
several subparts, each of which corresponds either to $S$ or to 
$S^*$, Eq. (\ref{start}), and the interpretation of them in physical 
terms is straightforward. 

Finally it is worth mentioning that the calculational rules of 
evaluating absorptive part of a generic thermal amplitudes are 
settled in Ref. 12). Finite-temperature generalizations of cutting 
rules are discussed in Refs. 12) and 13). 
\setcounter{equation}{0}
\setcounter{section}{3}
\section{Hard-thermal-loop resummation scheme in hot QCD 
\cite{sama,pis}} 
\def\theequation{\mbox{\arabic{section}.\arabic{equation}}}
When formally higher order correction to an (one-particle 
irreducible) amplitude is of the same order of magnitude as the 
lowest-order counterpart, a resummation of the \lq\lq correction'' 
is necessary. This is the case for classes of amplitudes whose all 
external momenta are soft, $P^\mu = O (g T)$. The relevant diagrams 
are the one-loop diagrams with hard loop momentum, 
$Q_{\mbox{\scriptsize{loop}}}^\mu = O (T)$, so is named the 
hard-thermal loop (HTL). 

The computation of 2-point amplitudes or the self-energy parts has 
been carried out long ago. Let us summarize the result. 

{\em Gluon:} In a covariant gauge, the full gluon propagator may be 
decomposed as \cite{land} 
\begin{eqnarray} 
& & \Delta_F^{' \, \mu \nu} (P) = 
   - {\cal P}_T^{\mu \nu} \, \Delta^{' \, T}_F (P) 
   - {\cal P}_L^{\mu \nu} \, \Delta^{' \, L}_F (P) 
   - \frac{1}{\lambda} \frac{D^{\mu \nu}}{P^2 + i 0^+}  
   - c (P) \frac{C^{\mu \nu}}{P^2 + i 0^+} \, , 
\label{decomp1} \\ 
& & \mbox{\hspace*{4ex}} \Delta^{' \, T}_F (P) = 
                            1 / (P^2 - \Pi_T (P)) \, , 
                            \;\;\;\;\;\;\;\; 
                         \Delta^{' \, L}_F (P) =  
                            1 / (P^2 - \Pi_L (P)) \, , 
\label{g-decomp} 
\end{eqnarray} 
where $\lambda$ is the gauge parameter and ${\cal P}_T$ 
[${\cal P}_L$] is the projection operator onto the transverse 
[longitudinal] or chromomagnetic [chromoelectric] sector. The third 
term on the rhs of Eq. (\ref{decomp1}) is the gauge term. Explicit 
form of ${\cal P}_{T / L}^{\mu \nu}$, $D^{\mu \nu}$ and 
$C^{\mu \nu}$ is given, e.g., in Ref. 6). 

The HTL contribution reads 
\begin{eqnarray} 
& & \Pi_T (P) = 
   - \frac{3}{2} m_g^2 
       \left[ \frac{p_0 P^2}{2 p^3} \, \ln 
           \left( 
                \frac{p_0 + p}{p_0 - p} 
           \right) - \frac{p_0^2}{p^2} 
       \right] 
            \, , 
\nonumber \\ 
& & \Pi_L (P) = 
   \frac{3}{2} m_g^2 \frac{P^2}{p^2} 
       \left[ 
            \frac{p_0}{p} \ln 
               \left( 
                  \frac{p_0 + p}{p_0 - p} 
               \right) 
            - 2 
        \right] 
          \, , 
\nonumber \\ 
& & c(P) = 
   0 \, , 
\nonumber \\ 
& & \mbox{\hspace*{3ex}} m_g^2 = 
   \frac{g^2}{3} \, T^2 
             \left[ 
             1 + \frac{N_f}{6} 
                \left( 1 + \frac{3}{\pi^2} \, \frac{\mu^2}{T^2} 
                \right) 
             \right] \, , 
\label{mass} 
\end{eqnarray} 
where $P$ is soft and $N_f$ is the number of quark flavors. We have 
assumed the common chemical potential $\mu$ for all $N_f$ 
(anti)quarks. Note that $\Pi_T (P)$ and $\Pi_L (P)$ are even 
functions of $p_0$. Observe that $\Pi (P)$'s are of $O (g^2 T^2)$, 
the same order of magnitude as the bare counterpart $P^2$. 

\hspace*{1ex} 

\begin{center} 
Characteristic features: 
\end{center} 

G1) 
{\em Landau damping}. Im $\ln [(p_0 + p) / (p_0 - p)] \neq 0$ for 
space-like $P^\mu$, $P^2 < 0$. 

G2) 
{\em Static limit}. Debye screening mass appears in the 
chromoelectric 
sector, $\Pi_L (p_0 = 0, \, p) = 3 m_g^2$. On the other hand, no 
screening mass appears in the chromomagnetic sector, $\Pi_T (p_0 = 
0, \, p) = 0$. The last fact indicates that, for some amplitudes 
that diverge (in naive perturbative calculation) due to the infrared 
singularity in the chromomagnetic sector, the screening is not 
sufficient for the amplitudes to converge. 

G3) 
{\em Dispersion curve} [the (positive) solutions, $p_0 = 
\omega_{T / L} (p)$, to $P^2 - \Pi_{T / L} (P) = 0$]. The mode with 
$P^2 - \Pi_L (P) = 0$, being absent in vacuum theory, is called the 
plasmon. $p_0 \geq m_g$: The solutions $\omega_{T / L} (p)$ exist 
for real $p$, showing the propagating modes. $\omega_{T / L} (p) - p 
> 0$, $\omega_{T / L} (p = 0) = m_g$ and, for $p >> m_g$, $\omega_T 
(p) \sim p + 3 m_g^2 / 4 p$ and $\omega_L (p) \sim p + 2 p \, e^{- 2 
p^2 / 3 m_g^2}$. The group velocities $v_{T / L} (p) \equiv d \, 
\omega_{T / L} (p) / d p$ are positive. $p_0 < m_g$: The solutions 
$p_0 = \omega_{T / L} (p)$ exist for pure imaginary $p$, showing the 
damping modes. 

{\em Quark:} The HTL-resummed soft-quark propagator takes the form, 
\begin{eqnarray} 
& & \displaystyle{ \raisebox{0.6ex}{\scriptsize{*}}} \! 
  S_F (P) = 
    - \frac{1}{2} 
       \left[ 
          \frac{\gamma^0 - \vec{\gamma} \cdot {\bf p} / p}{D_+ (P)} 
          + \frac{\gamma^0 + \vec{\gamma} \cdot {\bf p} / p}{D_- 
          (P)} 
       \right] 
           \, , 
           \label{q-moto} \\ 
& & D_{\pm} (P) = 
    - (p_0 \mp p) + \frac{m_q^2}{2 p} 
       \left[ 
            \left( 
               1 \mp \frac{p_0}{p} 
            \right) 
          \, \ln 
            \left( 
               \frac{p_0 + p}{p_0 - p} 
            \right) 
          \pm 2 
       \right] 
          \, , 
      \label{add} \\ 
& & \mbox{\hspace*{3ex}} 
    m_q^2 = 
      \frac{g^2}{6} \, T^2 
        \left( 1 + \frac{1}{\pi^2} \, \frac{\mu^2}{T^2} 
        \right) 
            \, . 
\end{eqnarray} 
Note that $D_+ (- p_0, p) = - D_- (p_0, p)$. Observe that the HTL 
contribution, the second term on the rhs of Eq. (\ref{add}), is of 
$O (g T)$, the same order of magnitude as the bare counterpart 
$- p_0 \pm p$. The $2 \times 2$ matrix propagator is related to 
$\displaystyle{ \raisebox{0.6ex}{\scriptsize{*}}} \! \hat{S}_F (P)$ 
through $\hat{M}_- (P) \displaystyle{ 
\raisebox{0.6ex}{\scriptsize{*}}} \! \hat{S}_F (P) \hat{M}_- 
(P)$ (cf. \S~2.2), where $\displaystyle{ 
\raisebox{0.6ex}{\scriptsize{*}}} \! \hat{S}_F (P) = \mbox{diag} 
\left[ \displaystyle{ \raisebox{0.6ex}{\scriptsize{*}}} \! S_F (P), 
\, -  \left( \displaystyle{ \raisebox{0.6ex}{\scriptsize{*}}} \! S_F 
(P) \right)^* \right]$. Taking the complex conjugate, in obtaining 
the $(2, 2)$ component of $\displaystyle{ 
\raisebox{0.6ex}{\scriptsize{*}}} \! \hat{S}_F (P)$, does not apply 
to the Dirac matrices.  

\hspace*{1ex} 

\begin{center} 
Characteristic features: 
\end{center} 

Q1) 
{\em Landau damping} as in the case of gluon. 

Q2) 
{\em Static limit}. Debye-like screening mass, 
$D_\pm (p_0 = 0, \, p) = \pm [p + m_q^2 / p]$. 

Q3) 
{\em Dispersion curve} [the (positive) solutions, $p_0 = 
\omega_\pm (p)$, to $D_\pm (P)$ $= 0$]. The mode with $D_- (P) = 0$, 
which is absent in vacuum theory, is called the plasmino. Both modes 
are the propagating modes. $\omega_\pm (p) - p > 0$, $\omega_\pm 
(p = 0) = m_q$ and, for $p >> m_q$, $\omega_+ (p) \sim p + 
m_q^2 / p$ and $\omega_- (p) \sim p + 2 p \, e^{- 2 p^2 / m_q^2}$. 
The group velocity of the $+$ mode, $v_+ (p) \equiv d \, \omega_+ 
(p) / d p$, is positive. The group velocity $v_- (p)$ of the 
plasmino shows an odd behavior. At $p = 0$, $v_-$ is negative and, 
as $p$ increases, $v_- (p)$ increases across $v_- = 0$ and 
approaches $v_- (p) = 1$. Furthermore, at large $p$, the residue 
$Z_-$ of the plasmino pole damps exponentially, $Z_- \sim 2 
(p^2 / m_q^2) e^{- 2 p^2 / m_q^2}$. 

\hspace*{1ex} 

Let me summarize the prominent features of HTL amplitudes. 

1) 
The HTL contributions, i.e., the contributions that are of the same 
order of magnitude as the lowest-order counterparts, arise in an 
$N$-gluon amplitude ($N \geq 2$) and a quark--antiquark--$N$-gluon 
amplitude ($N \geq 0$). For an amplitude including external FP-ghost 
lines, which appears in a covariant gauge, the HTL contribution does 
not appear.  

2) 
The HTL amplitudes are gauge independent. 

3) 
Kinetic-theory approach leads to the same result. \cite{blai-ian} 

4) 
In vacuum massless QCD, no HTL has arisen. This is because the gauge 
(chiral) invariance of the theory protects a gluon (quark) from 
getting corrections to the mass. 

5) 
The free part of the Lagrangian is modified so that the modified one 
(cf. \S 5) yields the HTL-resummed amplitudes. This means, among 
others, that, for soft modes, the in-field basis in vacuum theory, 
which is taken as the basis of perturbation theory, is not the good 
basis. It should be noted that the in-fields are irreducible 
representations of Poincar\'e group, which is a symmetry group of 
vacuum theory. However thermal field theory does not enjoy the 
Poincar\'e symmetry, so that the above result is not unnatural at 
all. In this relation I refer to Ref. 16). 

6) 
Let $G$ be the exact amplitude with soft external momenta and $H$ be 
the HTL contribution to $G$. In contrast to the case of vacuum 
theory, $(G - H) / H = O (g)$. 

7) 
HTL $N$-point amplitudes satisfy the Ward-Takahashi relation in {\em 
a stronger sense} than in vacuum theory. 

\hspace*{1ex} 

From these properties, especially from the last one, one can 
construct the $N$-point HTL amplitude $H^{(N)}$ ($N \geq 3$) from 
$H^{(2)}$. 
\setcounter{equation}{0}
\setcounter{section}{4}
\section{Effective action \cite{tay,R-E,pis-eff}} 
\def\theequation{\mbox{\arabic{section}.\arabic{equation}}}
Having obtained HTL $N$-point functions, one can construct an 
effective action $\displaystyle{ \raisebox{0.6ex}{\scriptsize{*}}} 
\! {\cal S}$, which is a generating functional of HTL $N$-point 
amplitudes. $\displaystyle{ \raisebox{0.6ex}{\scriptsize{*}}} 
\! {\cal S}$ is the leading contribution to 
${\cal S}_{\mbox{\scriptsize{eff}}}$ defined by 
\[ 
e^{i {\cal S}_{\mbox{\scriptsize{eff}}}} \equiv 
   \int_{\mbox{\scriptsize{hard modes}}} {\cal D} \psi \, {\cal D} 
   \overline{\psi} \, {\cal D} A \, e^{i {\cal S}} \, , 
\] 
where ${\cal S}$ is the QCD action. Various forms for 
$\displaystyle{ \raisebox{0.6ex}{\scriptsize{*}}} \! {\cal S}$ are 
available, from which I reproduce here the one obtained in Ref. 19): 
\begin{eqnarray*} 
\displaystyle{ \raisebox{0.6ex}{\scriptsize{*}}} \! {\cal S} & = & 
   - \frac{3}{4} m_g^2 \int d^{\, 4} x \, F^{\mu \alpha}_a (x) \, 
   \langle \frac{Y^\alpha Y^\beta}{(Y \cdot D_g)^2_{a b}} \, 
   F^\beta_{b \mu} (x) \rangle 
   - m_q^2 \int d^{\, 4} x \, \overline{\psi} (x) \, \langle 
   \frac{Y^\mu}{i \, Y \cdot D_q} \, \gamma_\mu \psi (x) \rangle 
   \, , \nonumber \\ 
& & \langle f \rangle \equiv \int \frac{d \Omega_{\hat{{\bf k}}}}{4 
   \pi} \, f (K) \, , \nonumber \\ 
& & Y^\mu \equiv (1, \, {\bf k} / k) \, , 
\end{eqnarray*} 
where $D_{g / q}$ is the covariant derivative acting on the 
gluon/quark field. 

$\displaystyle{ \raisebox{0.6ex}{\scriptsize{*}}} \! {\cal S}$ is 
also deduced \cite{blai-ian-eff} from the kinetic-theory approach. 

Various properties of $\displaystyle{ 
\raisebox{0.6ex}{\scriptsize{*}}} \! {\cal S}$ have been disclosed. 
For interested readers, I refer to the literature. 

1) Equation of motion and its solution. \cite{S-1} 

2) Conserved quantities via Neother's theorem or other means. 
\cite{S-2}  

3) Similarity to Chern-Simons theory. \cite{R-E,S-3,S-33} 

4) Classical nature of $\displaystyle{ 
\raisebox{0.6ex}{\scriptsize{*}}} \! S$. \cite{S-33,S-4} 
\setcounter{equation}{0}
\setcounter{section}{5}
\section{Hard modes \cite{nie-hard} with $|P^2| \leq O (g^2 T^2)$} 
\def\theequation{\mbox{\arabic{section}.\arabic{equation}}}
For illustration of the point, let me cut out the portion from a HTL 
gluon $N$-point amplitude with a quark loop, 
\begin{eqnarray} 
S_{i j} (P + K) (-)^{j - 1} \gamma^\mu S_{j k} (P) \, . 
\label{mass-s1} 
\end{eqnarray} 
Here $S_{i j}$ is the $(i, j)$ component of the bare quark 
propagator constituting the HTL, so that $P$ is hard $\sim T$. $K$ 
is the momentum of an external gluon and is soft $\sim g T$. 
Consider, e.g., the sector $i = j =1$. Equation (\ref{mass-s1}) 
contains $\delta (P^2) / (K + P)^2$. When the external momentum $K$ 
is on the mass shell $k_0 = \pm k$, this term develops well-known 
mass singularity at ${\bf p} \cdot {\bf k} = \pm p k$, $1 / 
(K + P)^2 \propto 1 / (1 + {\bf p} \cdot {\bf k} / p k)$, which, 
upon integration over ${\bf p}$, reflects on the logarithmic 
divergence of the HTL amplitude. This is the well-known mass 
singularity, which appears when the momentum $(K + P)^\mu$ can 
kinematically reach the light cone, $(P + K)^2 = 0$, on which the 
{\em bare} propagator $1 / (K + P)^2$ diverges. This observation 
leads us to analyze hard propagators near the light cone. 

Let us analyze the one-loop self-energy parts, $P \to Q + (P - Q) 
\to P$, with $P$ the hard external momentum. Recall that, in the 
case of self-energy part with $P$ soft, the hard $Q$ region (HTL) 
had yielded the dominant contribution. The soft modes and the hard 
modes are \lq\lq different'' modes. By contrast, for hard $P$, $Q$ 
and/or $P - Q$ are hard. When $Q$ [$(P - Q)$] is hard, one should 
use the self-energy-part-resummed propagator for the $Q$ $(P -Q)$ 
line, the self-energy part which we are to evaluate. 
[As a matter of 
course, when $Q$ [($P - Q$)] is soft, the HTL-resummed effective 
propagator should be used for the $Q$ [($P - Q$)] line.] 
Thus, we 
are lead to compute the self-energy part in a self-consistent 
manner. 

Here I display the result of the calculation, which is valid to 
leading order at logarithmic accuracy $\ln 1 / g >> 1$. 

{\em Gluon:} The self-energy-part resummed hard-gluon propagators 
$\displaystyle{ \raisebox{1.1ex}{\scriptsize{$\diamond$}}} 
\mbox{\hspace{-0.33ex}} \Delta_F (P)$'s in the covariant gauge 
read, with obvious notations, (cf. Eqs. (\ref{decomp1}) and 
(\ref{g-decomp})): 
\begin{eqnarray} 
& & \displaystyle{ \raisebox{1.1ex}{\scriptsize{$\diamond$}}} 
   \mbox{\hspace{-0.33ex}} \Delta_F^T (P) \simeq \frac{\epsilon 
   (p_0)}{2 p} \, \frac{1}{p_0 - \epsilon (p_0) [p + 3 m_g^2 / 4 p] 
   + i \epsilon (p_0) \gamma_T} 
   \label{d-t} \\ 
& & \displaystyle{ \raisebox{1.1ex}{\scriptsize{$\diamond$}}} 
   \mbox{\hspace{-0.33ex}} \Delta_F^L (P) \simeq 
   \displaystyle{ \raisebox{1.1ex}{\scriptsize{$\diamond$}}} 
   \mbox{\hspace{-0.33ex}} \Delta_F^g (P) \simeq 
   \frac{1}{P^2 + i 0^+} 
   \label{d-l} \\ 
& & \displaystyle{ \raisebox{1.1ex}{\scriptsize{$\diamond$}}} 
   \mbox{\hspace{-0.33ex}} \Delta_F^D (P) \simeq 
   \frac{1}{\lambda} \, \frac{1}{P^2 + i 0^+} 
   \label{d-d} \\ 
& & \displaystyle{ \raisebox{1.1ex}{\scriptsize{$\diamond$}}} 
   \mbox{\hspace{-0.33ex}} \Delta_F^C (P) \simeq 0 \, , 
   \label{d-c} \\ 
& & \mbox{\hspace*{4ex}} 
  \gamma_T = 
     \frac{g^2}{4 \pi} N_c T \, \ln (g^{- 1}) 
        \left[ 
           1 + O 
              \left( 
                 \frac{\ln \ln g^{- 1}}{\ln g^{- 1}} 
              \right) 
        \right] + O (g^2 T) \, . 
  \label{g-t} 
\end{eqnarray} 
Here $\Delta_F^g (P)$ is the FP-ghost propagator. Above forms are 
valid in the following regions: Im $\Delta_F^T (P)$; $||p_0| - p| 
\leq O (g^2 T \ln g^{- 1})$, Re $\Delta_F^T (P)$, $\Delta_F^L (P)$, 
$\Delta_F^g (P)$, $\Delta_F^c (P)$; $O (g^3 T) < ||p_0| - ( p + 3 
m_g^2 / 4 p) | \leq O (g^2 T \ln g^{- 1})$. $2 \times 2$ matrix 
propagators $\hat{\Delta}$'s are related to $\Delta_F (P)$'s through 
$\hat{\Delta} = \hat{M}_+ \hat{\Delta}_F \hat{M}_+$. 

Let us see how the bare propagators are changed through resummation 
of the self-energy part. I take Im $\displaystyle{ 
\raisebox{1.1ex}{\scriptsize{$\diamond$}}} \mbox{\hspace{-0.33ex}} 
\Delta_F^T (P)$. The bare form Im $\Delta_F^T (P) = - (\pi/ 2 p) \, 
\delta (|p_0| - p)$ turns out to be the \lq\lq smeared'' function Im 
$\displaystyle{ \raisebox{1.1ex}{\scriptsize{$\diamond$}}} 
\mbox{\hspace{-0.33ex}} \Delta_F^T (P)$, which is peaked at $|p_0| = 
p + 3 m_g^2 / 4 p$ with width $\gamma_T$. Note that $m_g^2 / p = O 
(g^2 T)$ while $\gamma_T = O (g^2 T \ln g^{- 1})$, so that $\gamma_T 
>> m_g^2 / p$ at logarithmic accuracy. For Re $\displaystyle{ 
\raisebox{1.1ex}{\scriptsize{$\diamond$}}} \mbox{\hspace{-0.33ex}} 
\Delta_F^T (P)$, similar observation may be made. 

Equations (\ref{d-t})~-~(\ref{g-t}) show that, as in the case of 
soft modes, the 
in-field basis in vacuum theory is not adequate for the 
transverse-gluon mode of hard $P^\mu$ with $P^2 \simeq 0$. 

{\em Quark} ($\mu = 0$): The self-energy-part resummed $2 \times 2$ 
quark propagators reads  
\begin{eqnarray*} 
& & \displaystyle{ \raisebox{1.1ex}{\scriptsize{$\diamond$}}} 
   \mbox{\hspace{-0.33ex}} S_{j i} (P) \simeq 
   \sum_{\tau = \pm} 
   \hat{{P\kern-0.1em\raise0.3ex\llap{/}\kern0.15em\relax}}_\tau \, 
   \displaystyle{ \raisebox{1.1ex}{\scriptsize{$\diamond$}}} 
   \mbox{\hspace{-0.33ex}} \tilde{S}_{j i}^{(\tau)} (P) \;\;\;\;\;\; 
   (\hat{P}_\tau = (1, \tau \hat{{\bf p}})) \;\;\;\;\;\; 
     (j, i = 1, 2) 
   \\ 
& & Re \, \displaystyle{ \raisebox{1.1ex}{\scriptsize{$\diamond$}}} 
   \mbox{\hspace{-0.33ex}} \tilde{S}_{1 1}^{(\tau)} (P) = 
   - Re \, \displaystyle{ \raisebox{1.1ex}{\scriptsize{$\diamond$}}} 
   \mbox{\hspace{-0.33ex}} \tilde{S}_{2 2}^{(\tau)} (P) 
   \\ 
& & \mbox{\hspace*{12ex}} \simeq 
        \frac{1}{2} \frac{p_0 - \epsilon (p_0) 
        (p + m_q^2 / p)}{[p_0 - \epsilon (p_0) (p + m_q^2 / p)]^2 + 
        \gamma_q^2} 
        \\ 
& & Im \, \displaystyle{ \raisebox{1.1ex}{\scriptsize{$\diamond$}}} 
   \mbox{\hspace{-0.33ex}} \tilde{S}_{1 1}^{(\tau)} (P) = 
   Im \, \displaystyle{ \raisebox{1.1ex}{\scriptsize{$\diamond$}}} 
   \mbox{\hspace{-0.33ex}} \tilde{S}_{2 2}^{(\tau)} (P) 
   \\ 
& & \mbox{\hspace*{12.5ex}} \simeq - \pi \epsilon (p_0) 
      \left[ 
         \frac{1}{2} - n_F (p) 
      \right] 
        \displaystyle{ \raisebox{0.9ex}{\scriptsize{$\diamond$}}} 
        \mbox{\hspace{-0.33ex}} \rho_\tau (P) \, , 
        \\ 
& & \displaystyle{ \raisebox{1.1ex}{\scriptsize{$\diamond$}}} 
   \mbox{\hspace{-0.33ex}} \tilde{S}_{1 2 / 2 1}^{(\tau)} (P) 
   \simeq 
      - i \pi \epsilon (p_0) 
        \left[ 
           \theta (\mp p_0) - n_F (p) 
        \right] 
     \displaystyle{ \raisebox{0.9ex}{\scriptsize{$\diamond$}}} 
     \mbox{\hspace{-0.33ex}} \rho_\tau (P) \, , 
     \\ 
& & \mbox{\hspace*{4ex}} 
      \displaystyle{ \raisebox{0.9ex}{\scriptsize{$\diamond$}}} 
      \mbox{\hspace{-0.33ex}} \rho_\tau (P) = 
      \frac{1}{\pi} \frac{\gamma_q}{[p_0 - \epsilon (p_0) 
      (p + m_q^2 / p) + \gamma_q^2]} 
      \\ 
& & \mbox{\hspace*{4ex}} 
     \gamma_q = 
        \frac{g^2}{4 \pi} C_F T \, \ln (g^{- 1})  
           \left[ 
              1 + O 
              \left( 
                 \frac{\ln \ln g^{- 1}}{\ln g^{- 1}} 
              \right) 
           \right] 
         + O (g^2 T) \, , 
\end{eqnarray*} 
where the Keldish variant of the real-time contour has been used. 

Similar observation to that in the case of gluon may be made. 

Substitution of the self-energy-part-resummed propagators screen the 
above-men\-tioned mass singularities and renders divergent integral 
finite. 
\section{Application to the computation of physical \\ 
quantities} 
\def\theequation{\mbox{\arabic{section}.\arabic{equation}}}
\setcounter{equation}{0}
\setcounter{section}{7}
Various physical quantities like thermal reaction rates are 
classified as follows. 

a) Computation in naive perturbation theory yields a finite 
result. 

b) Naive perturbation theory leads to a diverging result due to 
infrared (IR) singularity, which turns out to be finite within the 
HTL resummation scheme.   

c1) Same as above b) but it still diverges due to the IR 
singularity. 

c2) Same as above c1) but the divergence is due to the mass 
singularity. 

Examples of b) are the rate of hard photon and of hard photon-pair 
productions, energy loss of a particle, damping rate of a particle 
at rest etc. A typical example of c1) is the damping rate of moving 
particle and an example of c2) is the soft-photon production rare. 
A particle at rest feels only chromoelectric field and the IR 
singularity present in the computation within the naive perturbation 
theory is screened by the Debye mass. On the contrary, a moving 
particle feels chromomagnetic field also and, due to the absence of 
chromomagnetic screening mass, the screening at the IR region is not 
sufficient to render the diverging integral finite (cf. item G2) in 
\S~4). 

Hot QCD ($m_{quark} << T$) has only two parameters $g$ and $T$. 
Then, there arises natural hierarchy of scales: $T$ (hard), $g T$ 
(soft), $g^2 T$ (super soft), ... . 

Naive perturbation theory is valid at the hard region. 
HTL-resummation scheme deals with the soft region. Noting that the 
quantities classified into a) above receive a little contribution 
from the IR region, one can say that such quantities detect the 
\lq\lq physics'' in the hard region. The quantities classified into 
b) detect the \lq\lq physics'' at the soft as well as hard regions. 
The quantities belonging to c1) detect the super-soft, soft and hard 
regions. The scheme that deals with super-soft region in a 
consistent manner is not settled yet. In view of the fact that 
$(G - H) / H = O (g)$ (cf. item 6) at the end of \S~4), settlement 
of this scheme is an urgent issue but there is still a long way to 
go toward the solution. 

On the other hand, mass-singularity issue seems to be relatively 
easy to resolve. 

Let me mention, in turn, the damping rate of a moving particle and 
the soft-photon production rate (cf. c1) and c2) above). 

{\em Damping rate: \cite{sama}} Within the HTL-resummation scheme, 
the rate 
diverges at the IR end. It is expected that, at the next-to-leading 
order, the self-energy part acquires screening mass of $O (g^2 T)$ 
or \lq\lq something'' which screens the IR singularity. If this is 
the case, the diverging factor $\ln (g T / 0^+)$ turns out to be 
$\ln (g T / O (g^2 T)) \simeq \ln g^{- 1}$. 

In hot QED, however, it is generally believed that, in any order of 
perturbation series, no magnetic mass is induced. \cite{bla-scalar} 
In the IR region, the Bloch-Nordsieck approximation, $\gamma^\mu \to 
u^\mu$ (with $u^\mu$ the four velocity), is known to work. Employing 
this approximation scheme, it has been shown \cite{tak} that a 
moving hard electron damps according to $\propto e^{- \alpha T t \ln 
(m_e t)}$ ($t$: time, $m_e = e T / 3$: the QED counterpart of Eq. 
(\ref{mass})). 

{\em Soft-photon production rate:} The dominant contribution to 
$g^{\mu \nu} \Pi_{\mu \nu}$ ($\Pi_{\mu \nu}$ the photon polarization 
tensor) comes from an one-loop diagram with soft loop momentum. 
\cite{baier} Since all the relevant momenta are soft, one should use 
HTL-resummed effective quark propagators and HTL-resummed 
photon-quark vertices. As seen at the beginning of \S~6, the HTL 
photon-quark vertex diverges logarithmically $\sim \ln (g T / 0^+)$, 
because the external photon momentum is on the mass shell. 

According to the general argument in \S 6, substitution of 
self-energy-part-re\-sum\-med hard-quark propagators 
$\displaystyle{ \raisebox{1.1ex}{\scriptsize{$\diamond$}}} 
\mbox{\hspace{-0.33ex}} S$'s for the bare ones make diverging result 
finite, $\ln (g T / 0^+) \to \ln (g T / O (g^2 T)) \simeq \ln 
g^{- 1}$. 

However, the above substitution violates the 
Ward-Takahashi relation, which indicates that there must be 
important vertex corrections. Lebedev and Smilga have shown 
\cite{leb-smi} that the corresponding diagrams are the (resummation 
of) ladder diagrams. This yields the additional contribution to the 
soft-photon production rate, which coincides with the above 
contribution to leading order at logarithmic accuracy. 
\cite{nie-scr} Whether or not this analysis can be generalized to a 
generic reaction rate or thermal amplitude that belongs to the 
category c2) above is an open question. 
\section{Beyond the hard-thermal-loop resummation \\ 
scheme} 
\def\theequation{\mbox{\arabic{section}.\arabic{equation}}}
\setcounter{equation}{0}
\setcounter{section}{8}
As mentioned in the last section (cf. also the item 6) at the end of 
\S~4), toward establishing a next-to-leading-order resummation 
scheme is still a long way. Here I simply enumerate, without 
comment, some of the work made toward this end. 

1) Next-to-leading order computation of (chromoelectric) Debye mass. 
\cite{reb} 

2) Next-to-leading order computation of plasmon frequency. 
\cite{sch} 

3) Next-to-leading order computation \cite{fle-sch} of the gluon 
vacuum polarization tensor $\Pi_{\mu \nu} (P)$ with soft $P$. 

4) Next-to-leading order correction to the dispersion laws (cf. 
items G3) and Q3) in \S 4). \cite{fle-reb1} 

5) Self-consistent determination of chromomagnetic mass. \cite{ale} 

6) Improved effective action. \cite{fle-reb} 

Although not directly related to the subject of this review, I 
enumerate the following important achievements in the field of 
thermal field theory.  

1) Hot QED and hot scalar QED. \cite{bla-scalar,kia} 

2) Calculational scheme of grand partition function or pressure. 
\cite{bra-pre} 

Finally I mention the extensions to the nonequilibrium thermal field 
theory. For dealing with systems that are quasiuniform near 
equilibrium or quasistationary, traditional approach uses 
\cite{chou,hu} the Keldish variant of real-time formalism. As far as 
the computation of reaction rates are concerned, almost all the 
machineries of equilibrium thermal field theory hold as they are. An 
important one that does not hold is Eq. (\ref{self}), which causes 
the appearance of pinch singularity in self-energy-part-inserted 
propagators. \cite{alt-sei} Two approaches are devoted to this 
issue. 1) It has been shown \cite{alt} that such singular 
contributions can be resummed (see also Ref. 5)). Application of 
this result to the hard-photon production rate is made. 
\cite{bai-this} 2) A renormalization theory constructed through 
renormalizing number densities is proposed. \cite{nie-ren} This 
theory is same in structure as the equilibrium thermal field theory, 
so that no pinch singularity appears. 

Thermo field dynamics as mentioned in \S 2 (cf. item 2) after Eq. 
(\ref{self})) is generalized to the nonequilibrium case. \cite{ume} 
Recall that the Bogoliubov matrix $\hat{M}_\pm$ in Eq. (\ref{TFD}) 
defines the quasiparticle fields $\varphi$'s. Now $\hat{M}_\pm$ and 
then the quasiparticle picture depend on space-time coordinates. 
From Eq. (\ref{TFD}), we see that the time derivative of $\varphi$'s 
receives two contributions, one coming from the time derivative of 
the original fields $\phi$'s is governed by the Hamiltonian and the 
other comes from the time derivative of $\hat{M}_\pm$, through which 
the thermal energy is introduced. Then, through renormalization 
procedure of propagators, time dependence of $\hat{M}_\pm$ or the 
number density is determined. The determining equation turns out to 
be a generalized Boltzmann equation. 
\section{Conclusion} 
\def\theequation{\mbox{\arabic{section}.\arabic{equation}}}
\setcounter{equation}{0}
\setcounter{section}{9}
The structure of perturbative hot QCD is far more complicated than 
the perturbative vacuum QCD. Naive perturbation scheme, which is 
formulated using {\em in-field} basis (in vacuum theory) in a Fock 
space, is valid only at the low level of the \lq\lq QCD mountain'' 
(the short wave-length or the hard region $\lambda \sim 1 / T$). At 
the high level of the mountain (the long wave-length or the soft 
region $\lambda \sim 1 / g T$), naive perturbation scheme breaks 
down, which means that the in-fields are not the good basis at this 
level. The perturbation scheme that works here is the 
HTL-resummation scheme. Again this scheme does not apply at yet 
higher level of the mountain (the longer wave-length region). 
Continuous efforts aiming at establishing the new resummation scheme 
that works at this level are making. The goal is, however, still far 
a way. 

There are \lq\lq ravines'' along the light cone (mass singularities) 
here and there in the QCD mountain. Techniques of pass over these 
ravines are not completely settled yet. 

Comprehensive analysis of rates of various reaction taking place in 
nonequilibrium system, as well as the development of the 
theoretical framework of nonequilibrium quantum-field theory 
{\em per se}, have begun. 
\section*{Acknowledgements}
This work was supported in part by the Grant-in-Aide for Scientific 
Research ((A)(1) (No. 08304024)) of the Ministry of Education, 
Science and Culture of Japan. 


\newpage 
\begin{thebibliography}{99}
\bibitem{sama} M.~Le~Bellac, {\it Thermal Field Theory} (Cambridge 
Univ. Press, Cambridge, 1996). 
\bibitem{kapu}
J.I.~Kapusta, {\it Finite-Temperature Field Theory} (Cambridge Univ. 
Press, Cambridge, 1989).
\bibitem{ume}
H.~Umezawa, {\it Advanced Field Theory: Micro, Macro, and Thermal 
Physics} (AIP, New York, 1993).
\bibitem{das}
A.~Das, {\em Finite Temperature Field Theory} (World Scientific, 
Singapore, 1997). 
\bibitem{chou}
K.-C.~Chou, Z.-B.~Su, B.-L.~Hao and L.~Yu, 
Phys. Rep. {\bf 118} (1985), 1. 
\bibitem{land}
N.P.~Landsman and Ch.G.~van~Weert, 
Phys. Rep. {\bf 145} (1987), 141. 
\bibitem{henn}
P.A.~Henning, 
Phys. Rep. {\bf 253} (1995), 235. 
\bibitem{smi}
A.V.~Smilga, Surv.~High~Energy~Phys. {\bf 10} (1997), 233. [{\it 
Lecture at the 24 ITEP Winter School of Physics} (Moscow, Feb. 
1996).] 
\bibitem{nie}
A.~Ni\'egawa, Phys. Rev. D {\bf 40} (1989), 1199. 
\bibitem{nie1}
A.~Ni\'egawa, Phys. Lett. {\bf B247} (1990) ,351. \\ 
N.~Ashida, H. Nakkagawa, A.~Ni\'egawa and H. Yokota,   
        Phys. Rev. D {\bf 45} (1992), 2066; 
        Ann. Phys. (NY) {\bf 215} (1992) ,315; {\bf 230} (1994), 
        161(E). \\ 
Ashida~N., Int. J. Mod. Phys. {\bf A8} (1993), 1729. \\ 
A.~Ni\'egawa and K.~Takashiba,  
        Ann. Phys. (NY) {\bf 226} (1993), 293; {\bf 230} (1994), 
        162(E). \\ 
P.V.~Landshoff, Phys. Lett. {\bf B386} (996), 291. 
\bibitem{nie11}
A.~Ni\'egawa, Osaka City Univ. preprint OCU-PHYS 165 (1997). 
\bibitem{kobes} R.L.~Kobes and G.W.~Semenoff, 
           Nucl. Phys. {\bf B260} (1985), 714; {\bf B272} (1986), 
           329. 
\bibitem{cutting} 
S. Jeon, Phys. Rev. D {\bf 47} (1993), 4586. \\ 
P. F. Bedaque, A. Das, and S. Naik, hep-ph/9603325, 
MIT-CTP-2490, UR-1447, MRI-Phys-95-26. \\ 
F. Gelis, hep-ph/9701410, ENSLAPP-A-639/97. 
\bibitem{pis}
R.D.~Pisarski, Phys. Rev. Lett. {\bf 63} (1989), 1129. \\ 
E.~Braaten and R.D.~Pisarski, Nucl. Phys. {\bf B337} (1990), 569; 
{\bf B339} (1990), 310. \\ 
J.~Frenkel and J.C.~Taylor, Nucl. Phys. {\bf B334} (1990), 199. \\ 
H.~Vija and M.H.~Thoma, Phys. Lett. {\bf B342} (1995), 212. 
\bibitem{blai-ian}
J.-P.~Blaizot and E. Iancu, Nucl. Phys. {\bf B390} (1993), 589; 
{\bf B417} (1994), 608.
\bibitem{landsman} 
N.P.~Landsman, Ann. Phys. (NY) {\bf 186} (1988), 141. 
\bibitem{tay}
J.C.~Taylor and S.M.H.~Wong, Nucl. Phys. {\bf B346} (1990), 115. \\ 
J.~Frenkel and J.C.~Taylor, Nucl. Phys. {\bf B374} (1992), 156. 
\bibitem{R-E} R.~Efraty and V.P.~Naier, Phys. Rev. Lett. {\bf 68} 
(1992), 2891; 
Phys. Rev. D {\bf 47} (1993), 5601. 
\bibitem{pis-eff}
E.~Braatewn and R.D.~Pisarski, Phys. Rev. D {\bf 45} (1992), 1827. 
\bibitem{blai-ian-eff}
J.-P.~Blaizot and E.~Iancu, Nucl. Phys. {\bf B417} (1994), 608. 
\bibitem{S-1}
J.-P.~Blaizot and E. Iancu, Phys. Lett. {\bf B326} (1994), 138; 
Phys. Rev. Lett. {\bf 72} (1994), 3317. 
\\ 
R.~Jackiw, Q.~Liu and C.~Lucchesi, Phys. Rev. D {\bf 49} (1994), 
6787. 
\bibitem{S-2}
H.A.~Weldon, Can.~J.~Phys. {\bf 71} (1993), 300. \\ 
V.P.~Naier, Phys. Rev. D {\bf 48} (1993), 3432. \\ 
E.T.~Brandt, J.~Frenkel and J.C.~Taylor, Nucl. Phys. {\bf B410} 
(1993), 3. \\ 
J.-P.~Blaizot and E. Iancu, Nucl. Phys. {\bf B421} (1994), 565. \\ 
N.-P.~Chang, Phys. Rev. D {\bf 50} (1994), 5403; {\bf 51} (1995), 
4512. \\ 
See also, Y.~Hisamatsu and A.~Ni\'egawa, Phys. Rev. D {\bf 53} 
(1996), 3406. 
\bibitem{S-3}
R.~Jakiew and V.P.~Naier, Phys. Rev. D {\bf 48} (1993), 4991. \\ 
V.P.~Naier, Phys. Rev. D {\bf 50} (1994), 4201. 
\bibitem{S-33} 
J.-P.~Blaizot and E. Iancu, Nucl. Phys. {\bf B434} (1995), 662. 
\bibitem{S-4} 
P.F.~Kelly, Q.~Liu, C.~Lucchesi and C.~Manuel, Phys. Rev. Lett. 
{\bf 72} (1994), 3461. 
\bibitem{nie-hard} 
A.~Ni\'egawa, 
        Phys. Rev. D {\bf 55} (1997), 4997; {\bf 56} (1997), 
        2475(E). 
\bibitem{bla-scalar}
J.-P.~Blaizot, E.~Iancu and R.R.~Parwani, Phys. Rev. D {\bf 52} 
(1995), 2543. 
\bibitem{tak} 
J.-P.~Blaizot and E. Iancu, Phys. Rev. D {\bf 55} (1997), 973. \\ 
See also, K.~Takashiba, Int. J. Mod. Phys. {\bf A11} (1996), 2309. 
\bibitem{baier} 
R.~Baier, S.~Peign\'{e} and D.~Schiff, Z.~Phys. {\bf C62} (1994), 
337. \\ 
P.~Aurenche, T.~Becherrawy and E.~Petitgirard, 
Preprint ENSLAPP-A-452/93, NSF-ITP-93-155 (December, 1993). 
\bibitem{leb-smi}
V.V.~Lebedev and A.V.~Smilga, Physica {\bf A181} (1992), 187. 
\bibitem{nie-scr}
A.~Ni\'egawa, Phys. Rev. D {\bf 56} (1997), 1073. 
\bibitem{reb}
A.K.~Rebhan, Nucl. Phys. {\bf B430} (1994), 319. \\ 
P. Arnold and L. G. Yaffe, Phys. Rev. D {\bf 52} (1995), 7208. 
\bibitem{sch}
H.~Schulz, Nucl. Phys. {\bf B413} (1994), 353. 
\bibitem{fle-sch}
F.~Flechsig and H.~Schulz, Phys. Lett. {\bf B349} (1995), 504. 
\bibitem{fle-reb1}
F.~Flechsig, A.K.~Rebhan and H.~Schulz, Phys. Rev. D {\bf 52} 
(1995), 2994. 
\bibitem{ale}
G.~Alexanian and V.P.~Naier, Phys. Lett. {\bf B352} (1995), 435. 
\bibitem{fle-reb}
F.~Flechsig and A.K.~Rebhan, Nucl. Phys. {\bf B464} (1996), 279. 
\bibitem{kia}
U.~Kraemmer, A.K.~Rebhan and H.~Schulz, Ann. Phys. (NY) {\bf 238} 
(1995), 286. 
\bibitem{bra-pre}
E.~Bratten, Phys. Rev. Lett. {\bf 74} (1995), 2164. \\ 
E.~Braaten and A.~Nieto, Phys. Rev. D {\bf 53} (1996), 3421. \\ 
I.T.~Drummond, R.R.~Horgan, P.V.~Landshoff and A.~Rebhan, Phys. 
Lett. {\bf B398} (1997), 326. 
\bibitem{hu}
E.~Calzetta and B.L.~Hu, Phys. Rev. D {\bf 37} (1988), 2878. 
\bibitem{alt-sei}
T.~Altherr and D.~Seibert, Phys. Lett. {\bf B333} (1994), 149. 
\bibitem{alt}
T.~Altherr, Phys. Lett. {\bf B341} (1995), 325. 
\bibitem{bai-this} 
R.~Baier, M.~Dirks, K.~Redlich and D.~Schiff, Phys. Rev. D {\bf 56} 
(1997), 2548 and M.~Dirks, this issue, p. ???. 
\bibitem{nie-ren} 
A.~Ni\'egawa, hep-th/9709140 [Osaka City Univ. preprint OCU-PHYS 166 
(1997)]. 
\end{thebibliography}
\end{document}